\documentclass[12pt,preprint]{aastex}

\usepackage[english]{babel}
\usepackage[T1]{fontenc}
\usepackage[latin1]{inputenc}
\usepackage{amssymb}
\usepackage{color}
\usepackage[figuresright]{rotating}
\usepackage{amsmath}
\usepackage{ifthen}
\usepackage{acronym}
\usepackage{array}
\usepackage{url}
\usepackage{times}
\usepackage{longtable}
\usepackage{ifthen}

\newcommand{\figurewidth} {0.45\textwidth}

\newcommand{\ha} {H$\alpha$}

\newcommand{\Min}{${}^{\prime}$}
\newcommand{\Sec}{${}^{\prime\prime}$}

\newcommand{\electron} {$\mathrm{\bar{e}}$}

\shorttitle{EMCCD faint flux imaging}
\shortauthors{O. Daigle et al.}

\begin{document}


\title{Extreme faint flux imaging with an EMCCD}


\author{Olivier Daigle\altaffilmark{1,2,5}, Claude Carignan\altaffilmark{1,4}, Jean-Luc Gach\altaffilmark{2}, Christian Guillaume\altaffilmark{3}, Simon Lessard\altaffilmark{5}, Charles-Anthony Fortin\altaffilmark{5} and S\'ebastien Blais-Ouellette\altaffilmark{5}}
\email{odaigle@astro.umontreal.ca}

\altaffiltext{1}{Laboratoire d'Astrophysique Exp\'erimentale, D\'epartement de Physique, Universit\'e de Montr\'eal, C.P. 6128, succ. centre-ville, Montr\'eal, Québec, Canada, \mbox{H3T 2B1}}
\altaffiltext{2}{Aix-Marseille Universit\'e -- CNRS -- Laboratoire d'Astrophysique de Marseille, Observatoire Astronomique de Marseille-Provence, Technop\^ole de Ch\^ateau-Gombert,  38, rue Fr\'ed\'eric Joliot-Curie, 13388 Marseille, France}
\altaffiltext{3}{CNRS -- Observatoire Astronomique de Marseille-Provence -- Observatoire de Haute-Provence, 04870 St-Michel l'observatoire, France}
\altaffiltext{4}{Observatoire d'Astrophysique de l'Université de Ouagadougou, BP 7021, Ouagadougou 03, Burkina Faso}
\altaffiltext{5}{Photon etc., 5155 Decelles Avenue, Pavillon J.A Bombardier, Montr\'eal, Qu\'ebec, Canada, \mbox{H3T 2B1}}






\begin{abstract}
An EMCCD camera, designed from the ground up for extreme faint flux imaging, is presented. CCCP, the CCD Controller for Counting Photons, has been integrated with a CCD97 EMCCD from e2v technologies into a scientific camera at the Laboratoire d'Astrophysique Exp\'erimentale (LAE), Universit\'e de Montr\'eal. This new camera achieves sub-electron read-out noise and very low Clock Induced Charge (CIC) levels, which are mandatory for extreme faint flux imaging. Data gathered with the camera suggests that through enhanced manufacturing processes, which would avoid traps from being created, and with the help of the clock shapes producible with CCCP, the CIC generated during the vertical transfer could be virtually suppressed. The camera has been characterized in laboratory and used on the Observatoire du Mont M\'egantic 1.6-m telescope. The performance of the camera is discussed and experimental data with the first scientific data are presented.
\end{abstract}





\section{Introduction}

Charge Coupled Devices (CCD) are extremely sensitive devices, capable of detecting nearly all the photons that reach them, at certain wavelengths. Their read-out noise, however, renders impossible the detection of single photons accumulated into its pixels. In order to translate their high sensitivity into high signal-to-noise ratio images, the CCDs must, under faint flux, be exposed for an extended period of time before being read-out. Under some circumstances, such as for time resolved high resolution spectroscopy or photometry, the exposure time is limited, rendering the CCD unsuited.

\begin{figure*}[tbp]
\begin{center}
\includegraphics[width=\figurewidth]{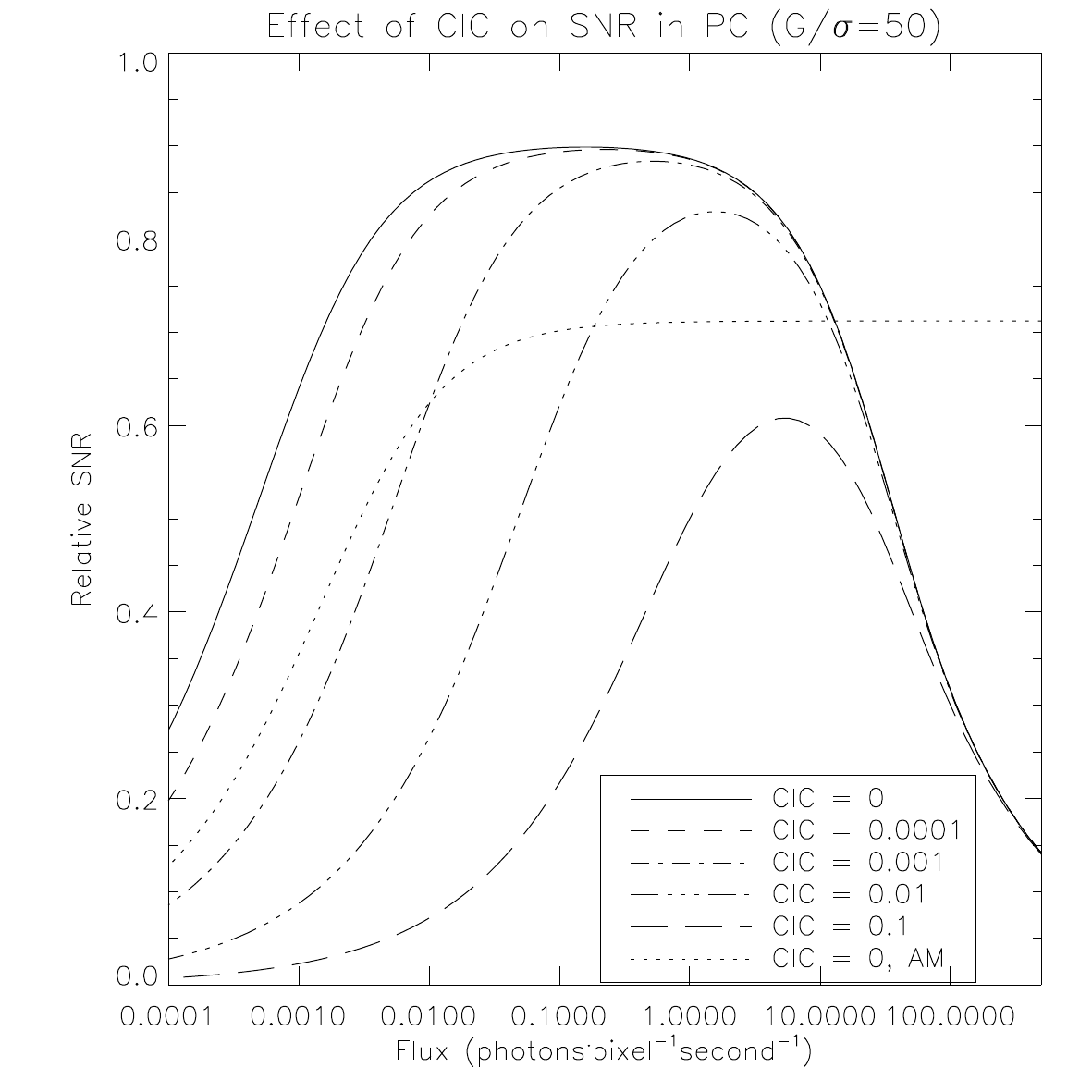}
\includegraphics[width=\figurewidth]{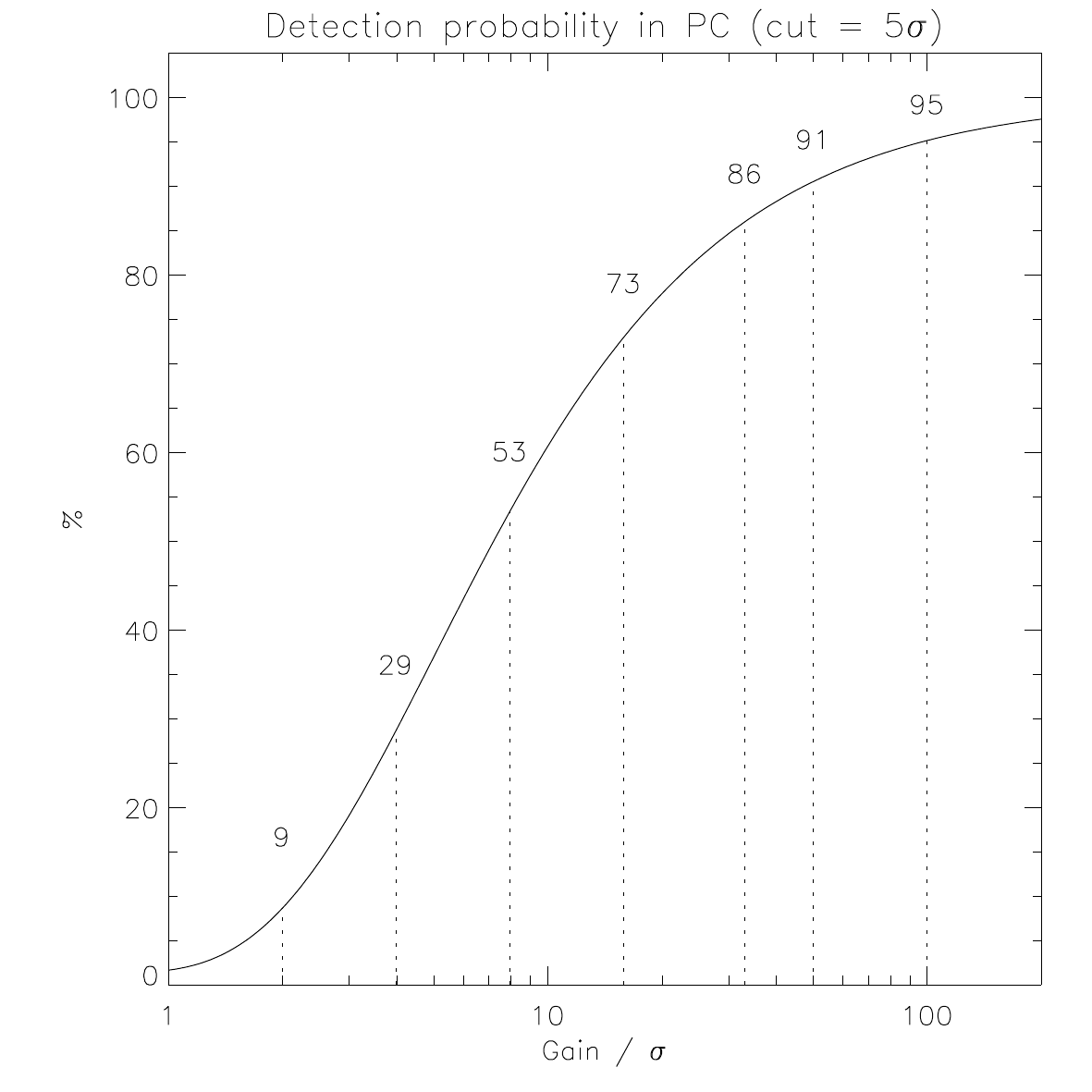}
\caption{\textbf{Left}: Effect of the CIC on the SNR of an observation. The SNR is plotted against the one of a perfect photon counting device, whose noise is only the shot noise, and that has the same Quantum Efficiency (QE). The simulation assumes a frame rate of 10 images/second, a dark rate of 0.001 \electron/pixel/second, and the CIC levels are expressed in \electron/pixel/frame. An extra plot shows the relative SNR of a perfect (no CIC) EMCCD operated in AM at the same frame rate. When there is no CIC, the frame rate has little effect on the SNR and this plot is a close approximation of the ultimate sensitivity of an EMCCD operated in AM. \textbf{Right}: The detection probability, in percent, of the photo-electrons as a function of the $G/\sigma$, for a single threshold of 5$\sigma$. Values for $G/\sigma$ ratios of 2, 4, 8, 16, 33, 50 and 100 are printed.}
\label{fig::cicPc}
\end{center}
\end{figure*}

The Electron Multiplying CCD (EMCCD) might come to the rescue \citep{2001SPIE.4306..178J}, with some intrinsic limitations. Whilst its EM register amplifies electrons before they reach the noisy charge-to-voltage amplifier, the multiplication process has a stochastic behaviour which results in an Excess Noise Factor (ENF). This ENF reaches a value of $\sqrt{2}$ at high gains \citep{stanford}. The ENF affects the observations processed in the so-called Analogic Mode (AM), where the amount of photons contained in a pixel is obtained by dividing the pixel's value by the EM gain. Besides, Clock Induced Charges (CIC), electrons that are generated as the device is read-out, completely dominate the noise budget when the device is operated at moderate and high frame rate. The physical processes involved in the generation of the CIC are outlined in \cite{techreport-minimal}.

One can get rid of the ENF by considering the EMCCD's pixel as being binary. This single threshold processing strategy, hereafter referenced as the Photon Counting (PC) strategy, simply considers the pixel as being empty if its output value is less than the threshold or as having accumulated one single photo-electron if its value is higher than the threshold. This processing have an obvious impact on the dynamic range of the EMCCD which can be compensated by taking short exposures and reading it often. In this regime, the CIC is the dominant source of noise and its impact on the Signal-to-Noise Ratio (SNR) is significant, even for a CIC rate as low as 0.01 \electron/pixel/frame (in \cite{2008AIPC..984..148T, 2008SPIE.7021E..10I, techreport-minimal}, CIC levels in the range of 0.01 -- 0.1 were typically measured) as shown by figure \ref{fig::cicPc}, left panel.

When operating the EMCCD in PC mode, the ratio of the EM gain over the real read-out noise (hereafter $G/\sigma$) sets the detection probability of a photo-electron (figure \ref{fig::cicPc}, right panel). For a given $\sigma$, a higher gain will have a higher probability of amplifying the photo-electron enough to take it above the threshold (typically set at 5 $\sigma$). Unfortunately, the EM gain can't be made arbitrarily high as the CIC scales with the EM gain \citep{2008SPIE.7014E.219D}.

Thus, the operation of an EMCCD in PC mode is challenging in many ways:
\begin{itemize}
\item The EMCCD must be operated at a high frame rate;
\item The CIC dominates over all the other sources of noise at high frame rate;
\item The $G/\sigma$ ratio must be high to allow the detection of a high proportion of photo-electrons;
\item The CIC scales with the EM gain;
\item The CIC is stronger in inverted mode, where holes populate the surface during the integration period;
\item It is preferable to try to operate the EMCCD in inverted mode to lower the impact of the dark signal: it would be useless to reduce the CIC at the price of a higher dark signal.
\end{itemize}

The last two points arise in part from the fact that Charge Transfer Efficiency (CTE) in the EM stage of the EMCCD suffers from low temperatures \citep{2008SPIE.7014E.219D}. In order to counterbalance this low CTE, higher clock amplitude must be used, resulting in an increased CIC (CIC is mostly independent of the operating temperature, for a given clock amplitude). It was found that an operating temperature of 183 to 188 K was a good compromise.

An easy conclusion one may draw from these facts is that the CIC must be taken to very low levels for the PC operation to be efficient (which is also expressed by figure \ref{fig::cicPc}, left panel). In order to do so, a whole new CCD controller was built. CCCP, the CCD Controller for Counting Photons, differs from the conventional controllers in the way it generates its clocks. Instead of being able to generate clocks with only 2 or 3 levels, the clock amplifiers of CCCP are driven by 12-bits DACs that are refreshed at a rate of \mbox{100 MHz}. This means that every 10 ns, any one of the $2^{12}$ levels of the clock can be sent to a clock input of the CCD. On the controller, 13 such arbitrary clocks are generated, together with a resonant high voltage clock (for the EM stage, having a switching precision of 1 ns). The controller also provides the biases and has a 16 bits high speed Analog-to-Digital Converter (ADC) coupled to an internal Correlated Double Sampler (CDS) and a digital communication interface to a computer for the transmission of the digitized images.

This controller was integrated into a LN$_2$-cooled cryogenic camera with a scientific grade CCD97 from e2v technologies. Using this camera, subsequently called CCCP/CCD97, it was possible to test and measure the effect of various clock shapes on the CIC. This paper presents these results and some discussions will be made around scientific results gathered with this camera.

\section{Effect on the vertical CIC}

\begin{figure*}[tbp]
\begin{center}
\includegraphics[width=\figurewidth]{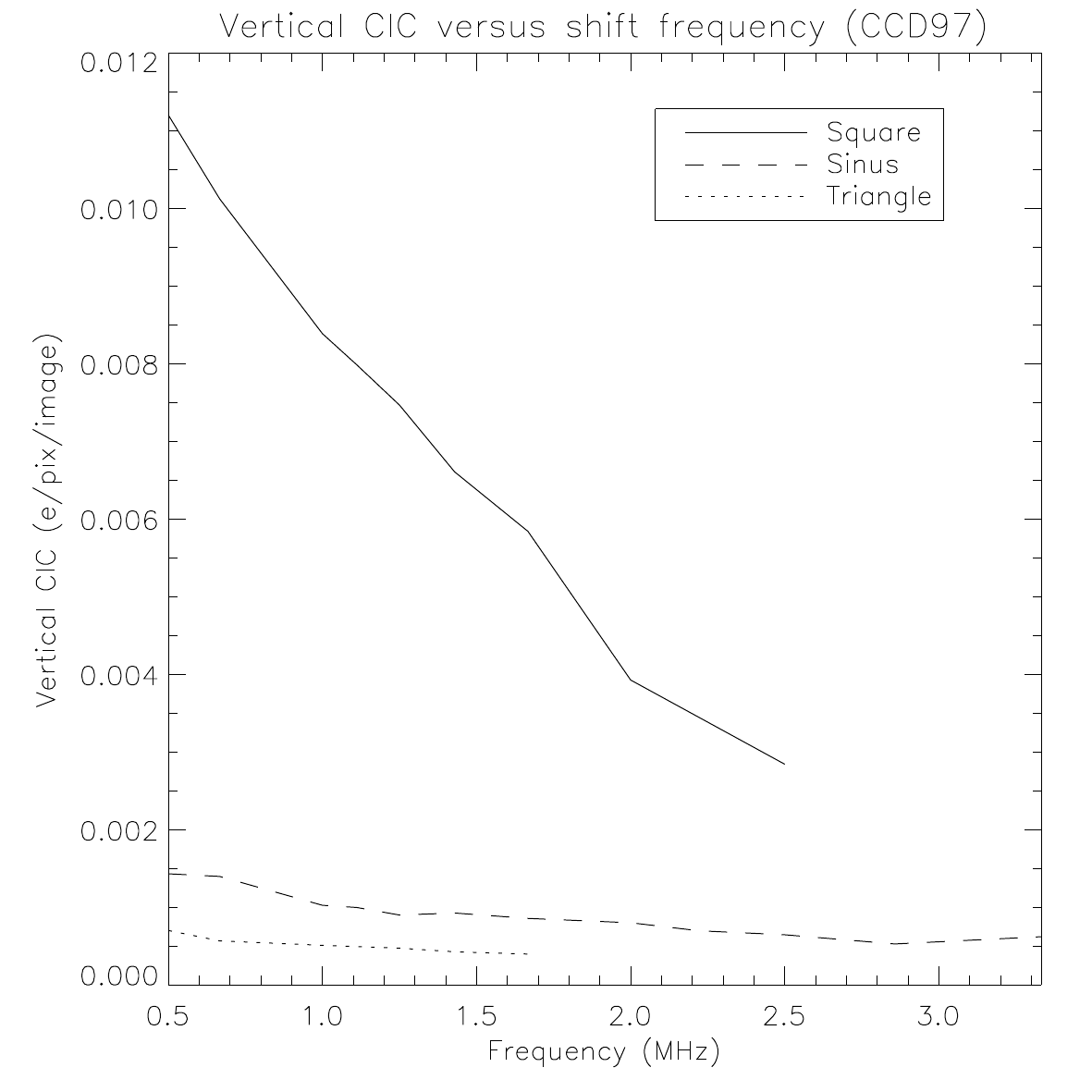}
\caption{Effect of various clock shapes on the CIC generated by the CCD97 during the vertical transfer of the charges as a function of the frequency of these clocks. For the square clock, the rise and fall times were \mbox{$\sim$150 ns} regardless of the frequency. For all clock shapes, the amplitude was set to the minimum (within $\sim$ 0.05 volt) at which the charge transfer occurred. This provided a full well about a half of the deepest well achievable on the device.}
\label{fig::verticalCIC}
\end{center}
\end{figure*}

The effect of the clock shape on the amount of generated CIC was measured. Results are presented in figure \ref{fig::verticalCIC}. This figure shows that the triangular waveform produce less CIC than its sinusoidal and square counterparts, for a given switching frequency. However, the sinusoidal waveform can move charges at a faster pace than the triangular clock. The minimum mean level of vertical CIC measured for the triangular clock is 0.0004 \electron/pixel/frame at \mbox{1.67 MHz}. The minimum mean level of vertical CIC measured for the sinusoidal clock is 0.0005 \electron/pixel/frame at \mbox{2.85 MHz}.

\begin{figure*}[tbp]
\begin{center}
\includegraphics[width=\figurewidth]{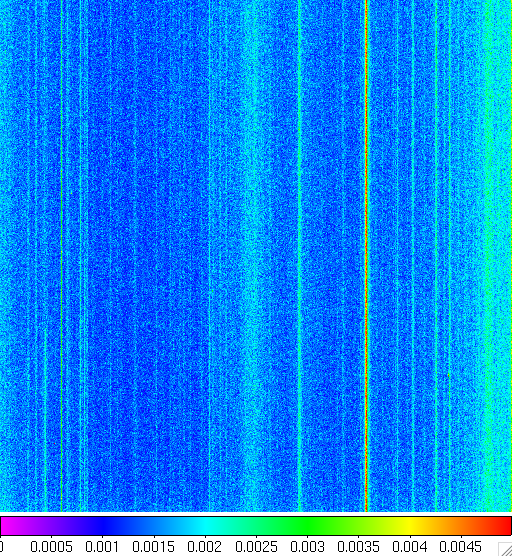}
\includegraphics[width=\figurewidth]{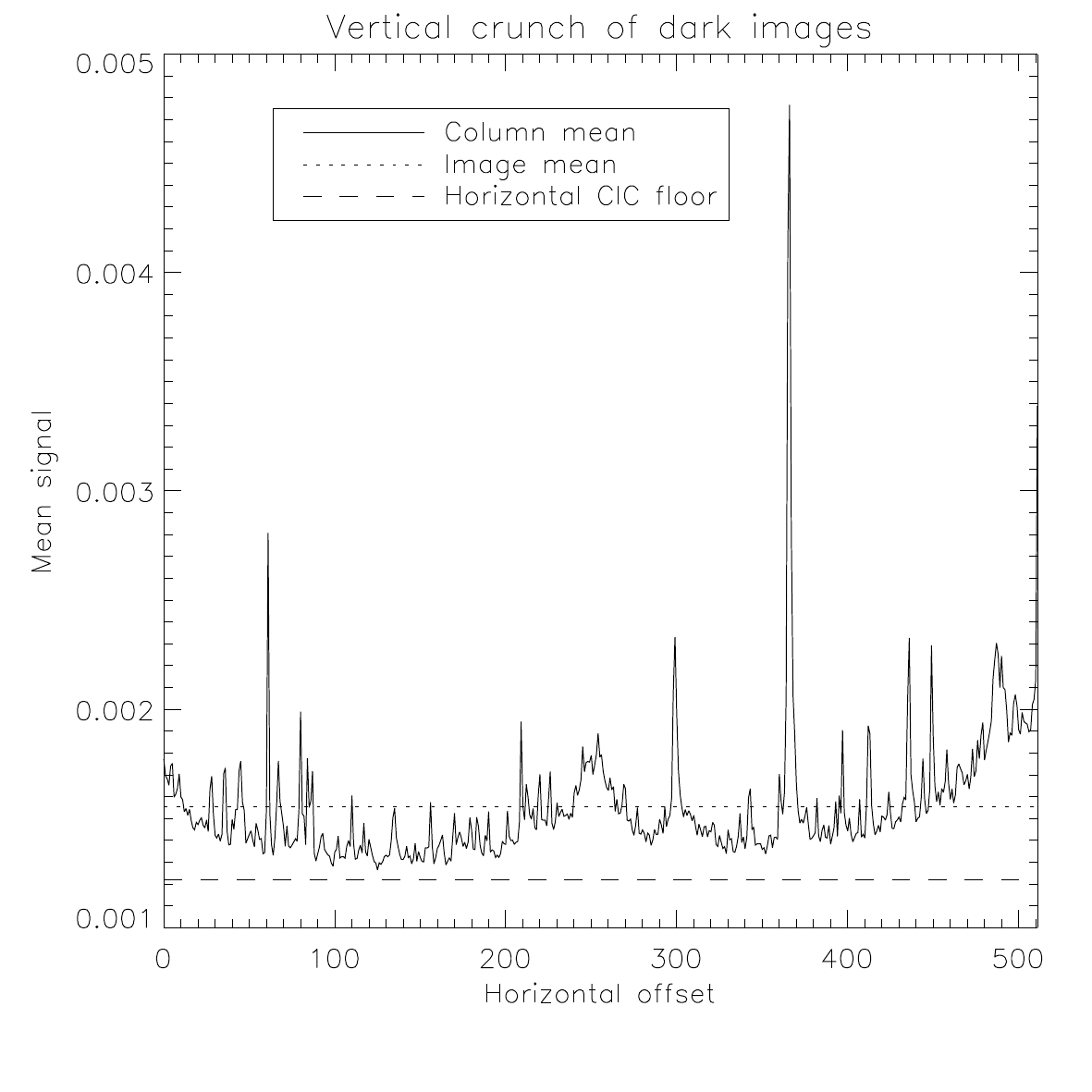}
\caption{\textbf{Left}: Mean of $\sim$100000 dark images of 0.05 second showing the structures created by the vertical CIC. \textbf{Right}: Mean signal level measured in every column of the dark images. The dashed line shows the amount of CIC that is attributable to the horizontal transfer. In some columns, there is little signal ($\sim$ 0.0001 \electron/pixel/image) above the horizontal CIC floor, which is constituted of an appreciable ($\sim$50\%) proportion of dark signal.}
\label{fig::verticalCICAnalysis}
\end{center}
\end{figure*}

Figure \ref{fig::verticalCICAnalysis}, left panel, shows the sum of about 100000 dark images of 0.05 seconds of integration, which are dominated by the CIC but not exempt of dark signal. These frames were acquired using sinusoidal vertical clocks. The vertical structures visible in this image clearly shows that the CIC generated during the vertical transfer is not homogeneous across the EMCCD. Some regions are strongly affected by the CIC: this is the case of the brightest column, which also exhibit a trap of $\sim$2 \electron. This trap is easily visible in flat fields taken under very low illumination. However, some regions of the image are a lot less affected by the vertical CIC. The right panel of this figure plots the mean signal level for every column of the image. The horizontal CIC floor plotted in this figure was obtained by overscanning the EMCCD in the horizontal direction and measuring the signal level of the overscanned region. This floor represents the mean mount of CIC and dark signal generated during the horizontal transfer of the charges. In some regions of the image (from column 90 to 200), very little signal is attributable to the vertical transfer. Furthermore, the dark signal level in these images is about 0.00004 \electron/pixel/image. When both the dark signal and the horizontal CIC are suppressed from the mean signal of these columns, one may conclude that these regions are nearly vertical CIC free. This suggests that through enhanced manufacturing processes, which would avoid traps from being created, and with the help of the clock shapes producible with CCCP, the CIC generated during the vertical transfer could be virtually suppressed, without resorting to the non-inverted mode operation.

\section{Effect on the SNR}

\begin{figure*}[tbp]
\begin{center}
\includegraphics[width=\figurewidth]{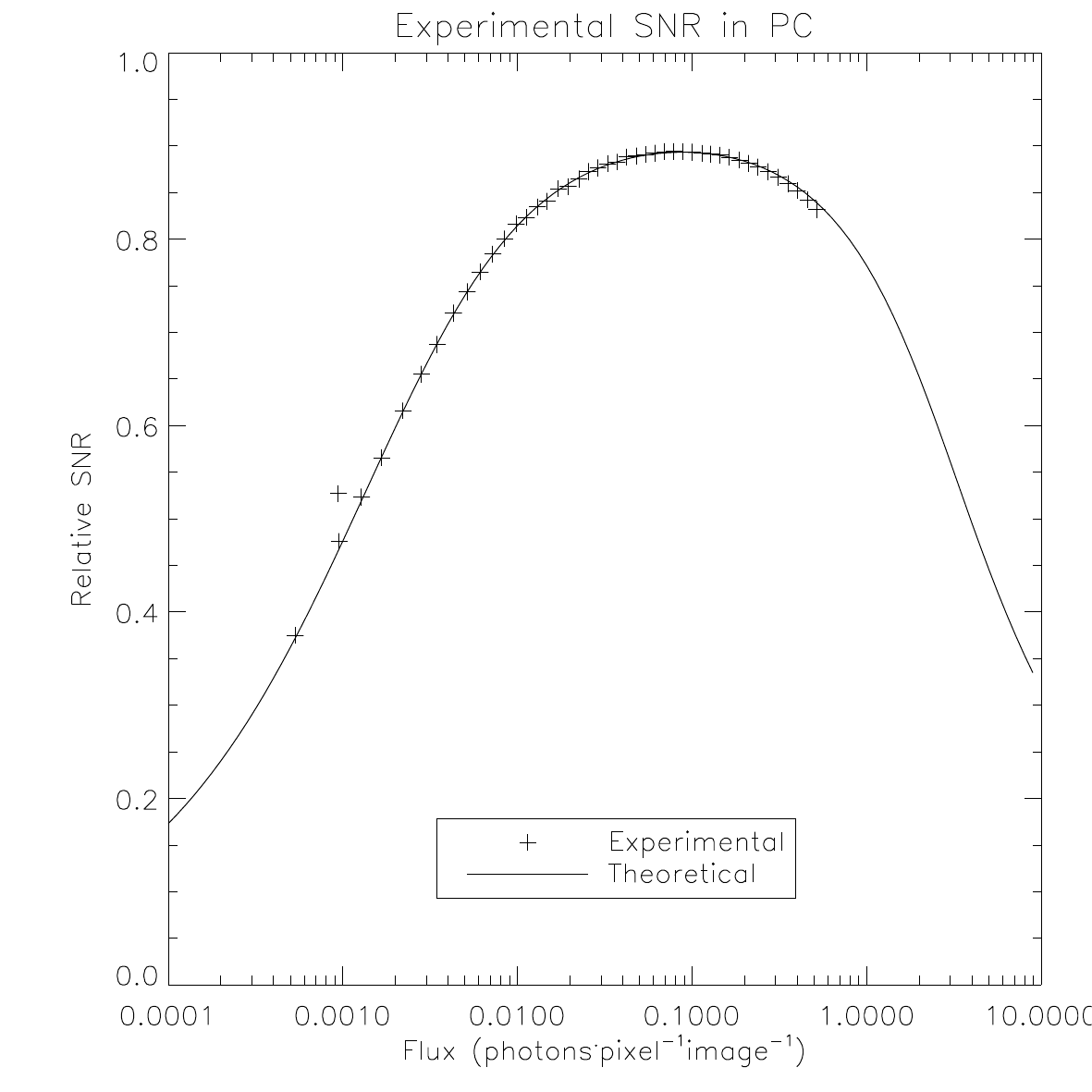}
\caption{Experimental SNR curve for the detected (QE and 5$\sigma$ threshold) photons. This curve was obtained by acquiring 80000 images with CCCP/CCD97 of a very low lit scene. The exposure time was 0.05s The acquisition was made at a $G/\sigma$ ratio of 33. The theoretical plot is the one of an EMCCD having the same $G/\sigma$ ratio and a CIC+dark rate of 0.0023. All data is plotted as the fraction of the SNR induced by the sole shot noise.}
\label{fig::experimentalSnrPc}
\end{center}
\end{figure*}

The SNR achieved with the camera, as a function of the incoming flux, was measured in laboratory. An extremely low lit scene, spanning about 3 orders of magnitude in flux, was observed, and the SNR achieved for a given flux range was calculated. Results are presented in figure \ref{fig::experimentalSnrPc}. The details of the measurements are explained in \cite{2009PASP..121..866D}. This figure shows that the CCCP/CCD97 camera provides 80\% of the SNR of a perfect photon counting device (for the same QE) for the 0.01 -- 1 photon/pixel/image flux range, at 20 frames per second. The peak SNR is attained at a flux of 0.09 photon/pixel/image and it is of 90\% the one of a perfect device.

\section{Scientific results}
The CCCP/CCD97 camera was installed on the 1.6-m telescope of the Observatoire du mont Mégantic (OMM). A focal reducer, which re-images the f/8 beam onto a f/2 focal plane was used, delivering a spatial resolution of 1.07\Sec\, per pixel and a diagonal field of view of 12\Min 55\Sec. This focal reducer has the advantage of providing an intermediate pupil where a Fabry-Perot (FP) interferometer can be placed, for integral field spectroscopy (IFS) applications. 

\subsection{Integral field spectroscopy}
The FP interferometer being a scanning instrument, it is of great interest to scan it multiple times throughout an observation to average the changing photometric conditions of the sky. Thus, these kind of observations requires many short exposures. The high ($>$ 10000) spectral resolution of the FP interferometer makes these observations read-out noise limited when using conventional CCD. This is the reason the FP interferometer was used with an GaAs Image Photon Counting System (IPCS) \citep{2002PASP..114.1043G, 2003SPIE.4841.1472H} for the recent observations (\cite{2008MNRAS.388..500E, 2008MNRAS.385..553D, 2006MNRAS.366..812C}, for example), and for an instrument lately developed for the William Hershel Telescope \citep{2008PASP..120..665H}. The IPCS has the great advantage of being read-out noise free and very lightly affected by the dark noise ($\sim 10^{-5}$ \electron/pixel/second). However, its QE is 28\% at maximum.

\begin{figure*}[tbp]
\begin{center}
\includegraphics[width=\textwidth]{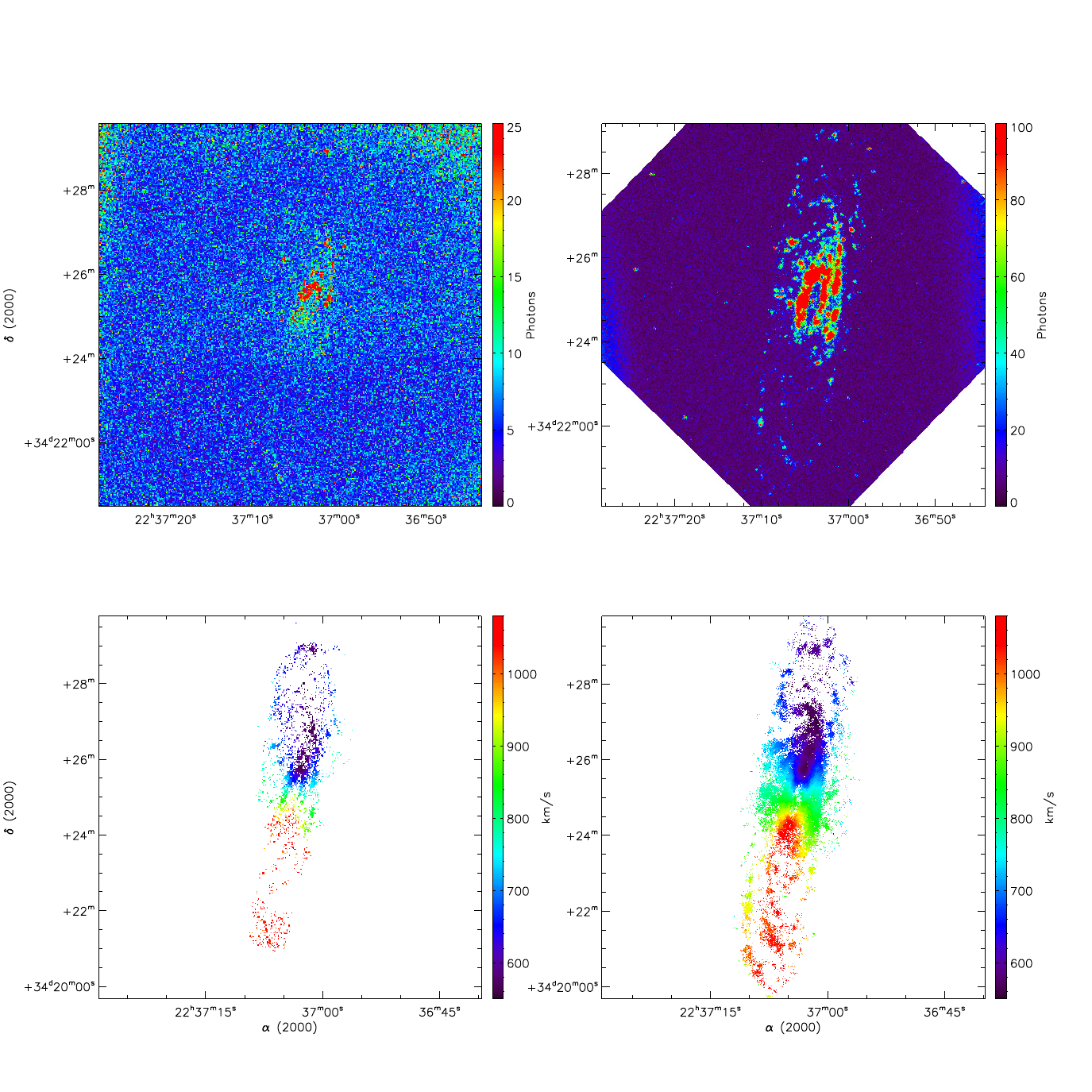}
\caption{IFS images of \mbox{NGC 7331} obtained with the IPCS (left) and with CCCP (right). The top images represents the monochromatic intensity of the \ha\, emission of the galaxy (continuum subtracted). The bottom images represent the velocity fields extracted from the IFS data. The observations were made in comparable sky conditions and the same total exposure time was used. The spectral resolution is about 15000.}
\label{fig::ngc7331}
\end{center}
\end{figure*}

The galaxy \mbox{NGC 7331} was first observed in \ha\, light with the FP and the IPCS in 2002 \citep{2006MNRAS.367..469D}. During an engineering run at the telescope in September 2008, this galaxy was re-observed with the CCCP/CCD97 camera. For this observation, the camera was operated at 2 frames per second. The results are presented in figure \ref{fig::ngc7331}. 

The gain in sensitivity achieved with CCCP/CCD97 as compared to the IPCS is obvious. Pixel for pixel, it was measured that the same SNR was achieved in 5 times less time with CCCP/CCD97 than with the IPCS. Thus, for the same integration time, the CCCP/CCD97 data reveals a wider, more precise, radial velocity field.

\section{Conclusions}
It has been demonstrated that careful shaping of the clocks of the EMCCD allows the CIC to be greatly reduced. In some regions of the CCD, the CIC is nearly completely suppressed. This suggests that enhancements in the manufacturing processes of the EMCCD, which would avoid traps from being created, could allow the production of EMCCDs that are a lot less affected by the CIC during the vertical transfer of the charges. 
By using these enhanced EMCCDs and the CCCP controller, one could expect this CIC component to be negligible as compared to the dark current, even at a high frame rate. Until this happens, however, the CCCP controller, which allow this shaping to be performed, renders possible the avoidance of the ENF by operating an EMCCD in PC mode at high frame rate. The SNR achieved by such a set-up is close to a perfect photon counting device, even for very low fluxes in the range of 0.01--1.0 photon/pixel/image. This camera was used at the OMM telescope for integral field spectroscopy applications at a high spectral resolution and has proven to be more sensitive than a GaAs IPCS for these applications. Readers interested in the details of the measurements presented in this paper are invited to read \cite{2009PASP..121..866D}.

\acknowledgments

O. Daigle is grateful to the NSERC for funding this study though its Ph. D. thesis. We would like to thank the staff at the Observatoire du mont M\'egantic for their helpful support.



{\it Facilities:} \facility{Observatoire du mont M\'egantic (OMM)}, \facility{Laboratoire d'Astrophysique Exp\'erimentale (LAE)}.





\clearpage



\end{document}